\documentclass[runningheads]{llncs}
\usepackage[T1]{fontenc}
\usepackage{hyperref}
\usepackage[caption=false,font=normalsize,labelfont=sf,textfont=sf]{subfig}
\usepackage{textcomp}
\usepackage{stfloats}
\usepackage{url}
\usepackage{verbatim}
\usepackage{graphicx}
\usepackage{cite}
\usepackage{xcolor}
\usepackage{soul}
\usepackage{lipsum}
\usepackage{enumerate}
\usepackage{blkarray}
\usepackage{multirow}%
\usepackage{amsmath,amssymb,amsfonts}%
\usepackage{mathrsfs}%
\usepackage{manyfoot}%
\usepackage{booktabs}%
\usepackage{listings}%

\begin{document}
\title{Multiscale Color Guided Attention Ensemble Classifier for Age-Related Macular Degeneration using Concurrent Fundus and Optical Coherence Tomography Images}
\titlerunning{Multiscale Color Guided Attention Ensemble Classifier (MCGAEc)}
%
\author{Pragya Gupta\inst{1} \and
Subhamoy Mandal\inst{2} \and
Debashree Guha\inst{2} \and Debjani Chakraborty\inst{1}}
\authorrunning{P. Gupta et al.}
%
\institute{Department of Mathematics, Indian Institute of Technology Kharagpur, West Bengal, 721302, India \\
\and
School of Medical Science and Technology, Indian Institute of Technology Kharagpur, West Bengal, 721302, India\\
}
\maketitle              
\begin{abstract} 
Automatic diagnosis techniques have evolved to identify age-related macular degeneration (AMD) by employing single modality Fundus images or optical coherence tomography (OCT). To classify ocular diseases, fundus and OCT images are the most crucial imaging modalities used in the clinical setting. Most deep learning-based techniques are established on a single imaging modality, which contemplates the ocular disorders to a specific extent and disregards other modality that comprises exhaustive information among distinct imaging modalities. This paper proposes a modality-specific multiscale color space embedding integrated with the attention mechanism based on transfer learning for classification (MCGAEc), which can efficiently extract the distinct modality information at various scales using the distinct color spaces. In this work, we first introduce the modality-specific multiscale color space encoder model, which includes diverse feature representations by integrating distinct characteristic color spaces on a multiscale into a unified framework. The extracted features from the prior encoder module are incorporated with the attention mechanism to extract the global features representation, which is integrated with the prior extracted features and transferred to the random forest classifier for the classification of AMD. To analyze the performance of the proposed MCGAEc method, a publicly available multi-modality dataset from Project Macula for AMD is utilized and compared with the existing models.      
\end{abstract}

\section{Introduction}{\label{sec 1}}
\raggedbottom
Diagnosing retinal disorders plays a vital role in guiding treatment decisions and improving outcomes for individuals with retinal conditions. Age-related maculopathy is a degenerative condition of the central region of the retina that is correlated with the cause of visual impairment that is recurring after 65 years of age \cite{bird1995international}. The diagnosis of AMD was first described in \cite{ryan1980disciform}. In the earlier phases of AMD, the patients have drusen and RPE abnormalities, whereas 
geographic atrophy and neovascularization of the retina may be interconnected with vision loss during the progression of the disorder. AMD can be characterized as dry or wet according to the pathogenesis. Choroidal neovascularization (CNV) \cite{green1986choroidal} is a manifestation of wet AMD and is diagnosed by analyzing the uncharacteristic expansion of blood vessels from the choroid into the retina. DME \cite{jemshi2018development} is a serious condition that can be attributed to hyperglycemia, which is a form of diabetic retinopathy (DR). It occurs due to prolonged exposure to high blood sugar, particularly in diabetic patients, which causes fluid leakage into the macula region, swelling, and thickening. Drusen \cite{auw2002optic} is a condition of dry AMD where tiny yellow or white deposits accumulate under the retina. If not characterized immediately, these ocular disorders can impair the retinal layer, especially the macular area, and perhaps end up with vision loss. Traditional diagnostic and grading systems for AMD are conducted by analyzing the color Fundus images \cite{ferris2013clinical}. Over the years, significant advancements have been made in medical diagnostics, offering new tools and techniques that enable precise and early detection of retinal disorders, especially the evolution of imaging techniques using Fundus images. OCT is one of the widely exploited diagnostic tools that will provide 3D structural information associated with the demonstration of cross-sectional images. OCT provides constructive information in investigating retinal disorders in challenging diagnostic cases and acquiring cross-sectional lesions of neovascularization associated with neighborhood tissue information. Clinical practitioners utilize OCT to examine the activity of AMD nowadays \cite{wilde2015diagnostic}. These techniques are effective; however, they suffer from intrinsic constraints, including specialized clinical practitioners and time-consuming, which induce variation in the ocular diagnosis, and hindered intervention can emerge.
\begin{figure}[htbp]
\centering
\begin{minipage}{0.3\textwidth}
    \centering
    \includegraphics[width=1.3in]{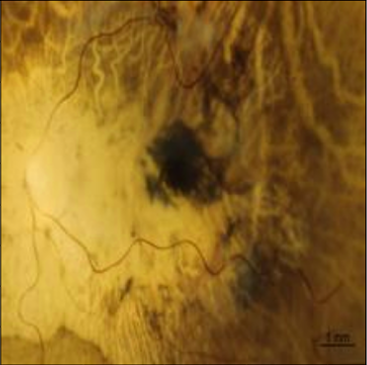}
    \\ {(a) }
\end{minipage}
\begin{minipage}{0.3\textwidth}
    \centering
    \includegraphics[width=1.3in]{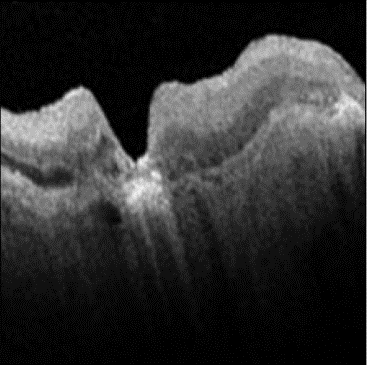}
    \\(b)
\end{minipage}
\caption{(a) The Original Fundus image and (b) the corresponding OCT image.}
\label{fig 1}
\end{figure}
Recent advancements in deep learning techniques have presented enormous possibilities for automated diagnosis tasks of retinal disorders at the expert level, decreasing the dependency on human experts in diverse fields \cite{zhang2023multi}. There is a rapidly growing interest in implementing deep learning models for classification using imaging modalities such as color fundus \cite{peng2019deepseenet} or OCT images \cite{karri2017transfer,das2020data}. Morano et al. \cite{morano2023weakly} proposed an explainable weakly-supervised technique for AMD diagnosis by operating on color fundus images. Philippe et al. \cite{burlina2019assessment} proposed to utilize Generative Adversarial Networks to obtain the synthetic image dataset for AMD classification. Tak et al. \cite{tak2021clinical} proposed a classification model based on a convolutional neural network (CNN) to categorize exudative and non-exudative classes. Researchers have introduced diverse types of deep learning models by extracting the distinct kind of feature characterization for distinguishing the abnormalities emerging in the fundus images for identifying the distinct stages of the AMD \mbox{\cite{el2023scale,ali2024amdnet23}}. However, the diagnosis of AMD by analyzing the fundus images and ignoring additional kinds of investigation is restricted due to the 2D representation of the fundus image. The fundus image is independent of slight changes in the macular breakage,  thickness, and detachment of the retinal layers \cite{saine2002ophthalmic}. On the other hand, OCT acquires a cross-sectional representation of biological tissues at microscopic spatial resolution \cite{van2007recent}, and it is a non-invasive technique. Liu et al. \cite{liu2011automated} proposed global feature image descriptors constructed based on machine learning for classifying four categories of AMD using OCT images. Karri et al. \cite{karri2017transfer} introduced a classification technique based on transfer learning using OCT images. Sun et al. \cite{sun2017fully} employed automatically align and crop of retina area followed by global illustrations by utilizing sparse coding, and finally, a multiscale support vector machine is executed for AMD classification. Although various automatic techniques have been introduced for the diagnosis of AMD and their different classes for examining the severity level based on OCT or fundus images. However, it is difficult to interpret intricate oculopathy with several lesions in the retina using a single imaging modality. Clinical practitioners generally consider two imaging modalities, as shown in Fig. \ref{fig 1}, including color fundus and OCT, in analyzing and diagnosing retinal disorders. Further, they simultaneously assess OCT and Fundus images and incorporate their specific feature representation details to provide accurate diagnoses. In this regard, several methods are introduced for the classification of ocular disorders using multi-modality imaging techniques \mbox{\cite{lam2024performance}}. Yoo et al. \mbox{\cite{yoo2019possibility}} employed a random forest classifier with a VGG model for the classification of AMD using multi-modality OCT and fundus images. Yin et al. \mbox{\cite{dai2021transmed}} proposed a TransMed method using multi-modality medical images for the classification tasks. Li et al. \mbox{\cite{li2021multi}} used multi-modal evidence and introduced a multi-instance deep learning model for the diagnosis of retinal disorders. Wang et al. \mbox{\cite{wang2022learning}} proposed a two-stream CNN model for the classification of ocular diseases using Fundus and OCT images. Fang et al. \mbox{\cite{fang2021multi}} proposed a technique by integrating the regression approach to the deep learning models for the diagnosis of glaucoma grading. Xing et al. \mbox{\cite{xing2022advit}} proposed a transformer-based model using mult-modality images. These introduced deep learning models based on multi-modality fundus and OCT images improve the performance of diagnosis of retinal disorders compared to the utilization of single modality imaging methods. Notably, relying on the single color channel and scale space of the multi-modality images employed for the deep learning models often ignores capturing the global and local feature representation in distinct multi-scale space. Moreover, the feature information in the distinct color spaces provides diverse characteristic feature information that is not attainable by employing a single color space.
\newline
In this work, we propose modality-specific multiscale color space embedding based on the attention mechanism for the classification of different stages of the AMD. The proposed framework includes the utilization of YCbCr and HSV color space for fundus images on different scale spaces, followed by OCT images at multiscale to capture the crucial features by exploiting the different characteristic feature representations in distinct color spaces and fed to three different pre-trained VGG16 model for the feature extraction. It enhances the generalization ability of the proposed model and also considers the different feature attributes of the color spaces by considering distinct scales into a unified framework. An attention mechanism is incorporated to extract the most representative feature from each considered path, followed by a concatenation of the extracted features from each pre-trained model, and the attention mechanism is transferred to a random forest classifier (RFC) model for the classification of AMD. To experimentally analyze the interpretation of the proposed MCGAEc method, a publicly available multi-modality dataset is utilized, which is provided
at Project Macula \cite{schaal2015outer} for AMD and compared with the existing method to show the significance of the proposed framework. In summary, the key contributions to this study are mentioned below:

\begin{itemize}
    \item  We propose a modality-specific multiscale color space embedding integrated with the attention mechanism, which adaptively specifies the most representative features and is fed to the RFC model for the classification of AMD.
    \item  To analyze the efficacy of the proposed MCGAEc model, experiments are performed over a publicly available dataset of the color fundus and OCT images (Project Macula) and compared with the single modality (Fundus or OCT) images on different color spaces at distinct scales.
    \item The proposed method is compared with the state-of-the-art (SOTA) method to demonstrate the efficacy of the proposed MCGAEc model using the evaluation measures.
\end{itemize}
The remainder of the manuscript is systematized as follows: Section \ref{sec 2} demonstrates the proposed framework, followed by experimental analysis in Section \ref{sec 3}. Section \ref{sec 4} includes the concluding observations.
\section{Proposed Model}
\begin{figure}[!t]
    \centering
    \includegraphics[width=5.5in]{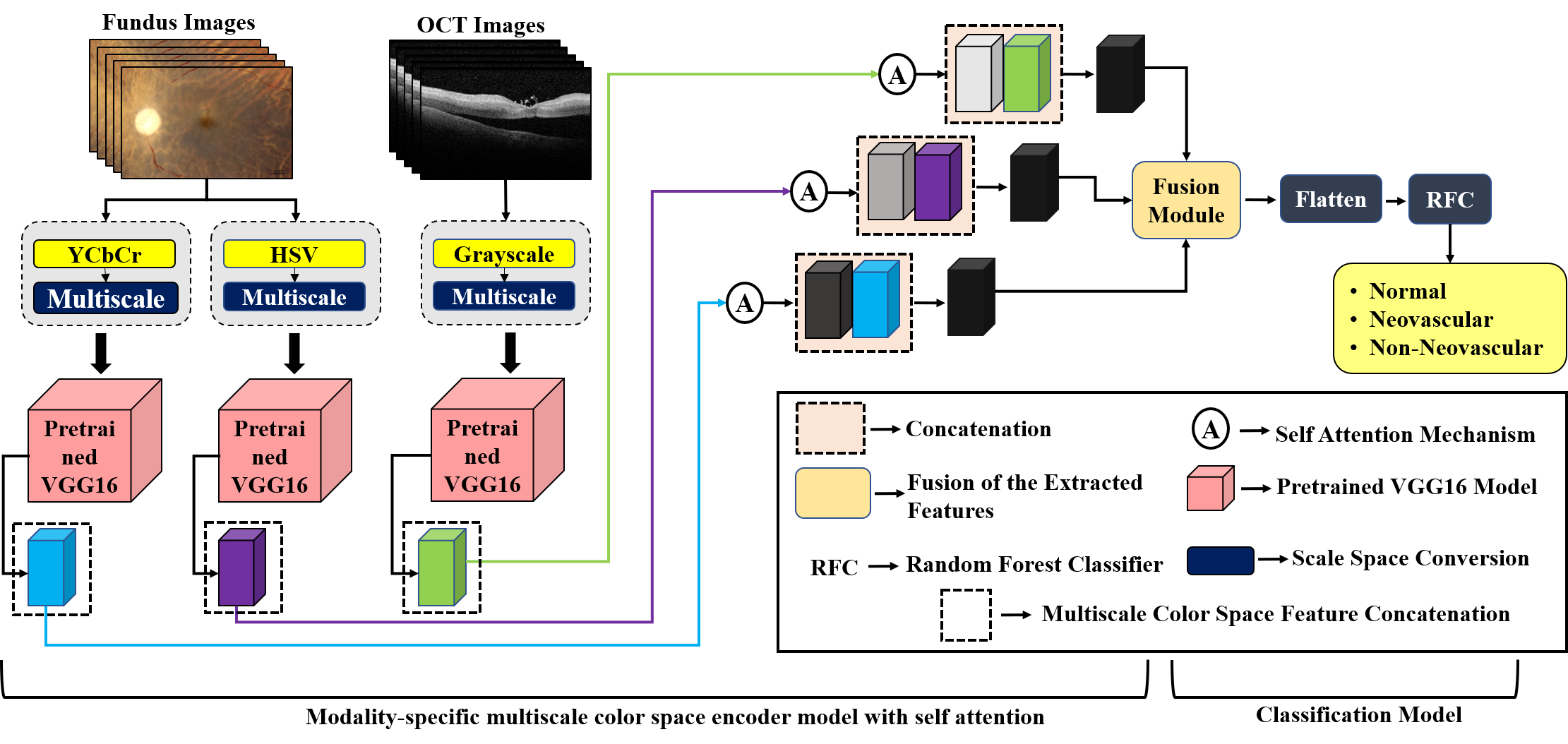}
    \caption{The proposed framework for the combination of the Fundus and OCT images for the classification of AMD}
    \label{fig 2}
\end{figure}

\begin{figure}[htbp]
\centering
\begin{minipage}{0.3\textwidth}
    \centering
    \includegraphics[width=1.0in]{Fig_1.png}
    \\ {(a) }
\end{minipage}
\hspace{-1.3cm}
\begin{minipage}{0.3\textwidth}
    \centering
    \includegraphics[width=1.0in]{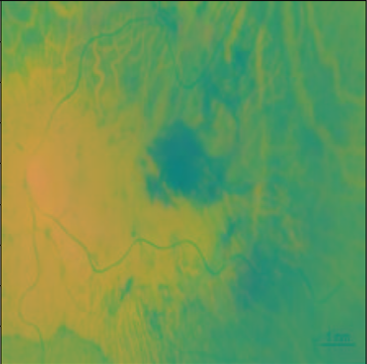}
    \\(b)
\end{minipage}
\hspace{-1.3cm}
\begin{minipage}{0.3\textwidth}
    \centering
    \includegraphics[width=1.0in]{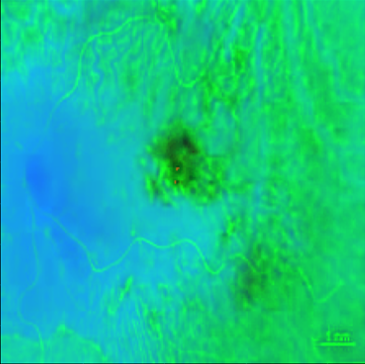}
    \\(c)
\end{minipage}
\caption{Representation of the Fundus image in (a) RGB color space (original), (b) YCbCr color space, and (c) HSV color space.}
\label{fig 3}
\end{figure}

{\label{sec 2}}
\raggedbottom
This section introduces the proposed MCGAEc framework illustrated in Fig. \ref{fig 2}. MCGAEc comprises a modality-specific multiscale color space encoder module in which a fundus image is transformed into distinct color spaces, say, YCbCr and HSV, which are transferred to two encoder paths. For OCT images, one encoder path named gray-scale is constituted. In each path, the transformed Fundus images are forwarded to multiscale space conversion, followed by input for the pre-trained VGG16 model to acquire vital features from each considered color space at various scales. The extracted features from each path from the Fundus images are transferred to the self-attention module. Simultaneously, we extract the features from the gray-scale path for the OCT images. Then, features extracted from the pre-trained VGG16 and self-attention module are concatenated. Finally, we fused all the extracted features from each path and fed them to the RFC for the classification of AMD.   
\textbf{Modality-Specific Multiscale Color Space Encoder Model}: The color variations of fundus images enclose exhaustive ranges, and heterogeneity in color casts restricts the classical models \cite{chandra2021survey}. Motivated by traditional enhancement techniques that function over different color spaces \cite{iqbal2010enhancing,naik2003hue,mandal2010segmentation}, we extract distinct characteristic features from two color spaces (HSV, YCbCr) where the identical fundus image has distinct pictorial representation in diverse color spaces demonstrated in Fig. \ref{fig 3}. The fundus image is explicitly to visualize in RGB color space because of its intense physical significance in color. However, the color segments R, G, and B are positively associated and are easy to be influenced by the variation of luminance, occlusion, and other factors. On the other hand, YCbCr color space can intuitively reminisce the luminance (Y) and two chroma components (Cb and Cr). YCbCr is crucial in digital images and video to separate luminance from its chrominance. This fragmentation is beneficial because the human eye is more susceptible to luminance than chrominance, and it entitles more efficiency in compression that aligns with the visual perception of humans. HSV color space characterizes the hue, saturation, contrast, and brightness of the Fundus image. The considered color space has diverse characteristics and benefits. To integrate their properties in the fundus image feature enhancement, we assimilate the characteristics of distinct color spaces into a unified deep characteristic model. Furthermore, the color variations of two considered points with a diminutive variation in one color space can be enormous in another color space. Thus, the distinct color space integration can facilitate the measure of the color divergence of fundus images. 

To extract the necessary features from distinct color spaces, which is required for the classification of AMD, multi-scaling is incorporated. When the scale is augmented, most of the noise is eliminated. If the features are available in more than one coarse scale, which indicates it should be available at different scales as well \cite{sumengen2005multi}. Therefore, this strategy is implemented over each color space by considering different scales for fundus and OCT images, respectively. Afterward, we executed the pre-trained VGG16 \cite{simonyan2014very}, which includes 16 deep layers for extracting the features from considered multiscale color space for fundus and OCT images. VGG16 is an adequate model for the image classification task, and the pre-trained model enables to extraction of features from an extensive corpus of images when utilizing a small dataset. VGG16 model provides the balance between the performance and computational efficiency compared to other complex models. Additionally, it allows the experiments without requiring comprehensive computational resources, which provides an advancement to integrate the attention layer with that.
A transfer learning mechanism is utilized, and extracted features from each are fed to the distinct attention module. 

\textbf{Self-Attention}: The self-attention mechanism was first introduced in the domain of image processing \cite{shaw2018self}, which is integrated into the attention layer. It enables to concentrate on salient or global features of the datasets. It provides an adequate correlation towards global feature information within each single image. The primary notion behind the self-attention mechanism is to associate weighted average values evaluated from the prior layers, and the attention weights are assessed as follows:

\begin{equation}{\label{eq 1}}
    Attenttion(Q, K, V)=softmax\Big(\dfrac{QK^{T}}{\sqrt{d_{head}}}\Big)V
\end{equation}
where Q is a query, K is the key, and V is the value. In the assertion of the precise characterization of each multiscale color space feature, the extracted features from the pre-trained model through each path have dissimilar contributions. Consequently, we utilize a self-attention mechanism to exploit the inter-variability between the extracted features obtained through each path. In the proposed framework, the extracted features from each path through the pre-trained VGG16 model are used as input for the self-attention module. The extracted features through the pre-trained VGG16 model are confined by structure to concentrate primarily on local features of fundus and OCT images while incompetent to acquire the global feature information.  Self-attention focuses on specific global features of the images. Therefore, the extracted features from the pre-trained deep network model are fed into the self-attention model corresponding to each path. The attention features and extracted prior features from the pre-trained model from each path are fused together by concatenation for a better representation of each feature acquired through the Fundus and OCT images, which are used as input for the classification model.

\textbf{Classification Model}: The computed features through the modality-specific multiscale color space encoder model with self-attention are transferred into the supervised machine learning RFC \cite{kulkarni2013random}. The RFC comprises an amalgamation of tree classifiers where a particular classifier is acquired through a random vector, which is sampled individually from the input data, and the respective tree provides a unit vote for the most prevalent class to categorize input data. The RFC is utilized for the classification of the input data, which relies on the utilization of randomly chosen features at each node to expand a tree. In RFC, features are illustrated on the inner nodes, which are called decision nodes, and enable to generate the predictions from a sequence of feature-based fragmentation. RFC employs a collection of random decision trees and integrates them concurrently to construct a robust model that is less acute to the training data. The RFC algorithm is capable of handling higher dimensional data and utilizing an enormous number of trees in the combination. The output of the RFC is computed by a majority counting of votes obtained through trees. Here, RFC is used, which includes considerably lower computational complexity as each particular tree exclusively utilizes a part of the input vector in a Random Forest. 
\section{Experimental Framework}{\label{sec 3}}
\raggedbottom
In this section, the description of the multi-modality OCT and fundus image datasets for diagnosis of AMD is provided, followed by evaluation measures to check the performance of the proposed MCGAEc model. Finally, a comprehensive empirical study, including the ablation analysis, is given to show the significance of the proposed MCGAEc model for the classification using the multi-modality dataset. 
\subsection{Dataset and Evaluation Measures}{\label{subsec 3.1}}
The experimental analysis of the MCGAEc model is assessed over the publicly open multi-modality OCT and fundus images dataset at the Project Macula \cite{schaal2015outer} for AMD classification (\url{https://projectmacula.cs.uab.edu}). The publicly available dataset aimed to investigate AMD in patients and their severity level. The diagnosis of AMD over the provided dataset is endorsed by pathohistological examination. The dataset is categorized into three classes: normal, non-neovascular, and neovascular. The normal class includes 50 OCT and Fundus images each, followed by 19 and 40 in non-neovascular and neovascular, respectively. 
\\
The spatial resolution of Fundus and OCT images in each class is of different variation, and data is not enough to train the model. Therefore, data augmentation is applied with rotations, translations, and contrast changes for increments in the number of images. Data augmentation \cite{shorten2019survey} is extensively applied to enhance the generalization of the proposed method. We erratically retrieved 500 fundus images for each class and 500 OCT images for each class that matched with the Fundus images. We have performed the rotation in a range $[-25^{\circ}, +25^{\circ}]$, with translation $[-10\%, +10\%]$ of the width of the image, and contrast change with ranges of $[-50\%,+50\%]$. All the generated OCT and fundus images are resized $224 \times 224$ for the input of the pre-trained model. The experiments are performed in the selection of an optimal number of trees from the set $\{100,300,500, 700, 1000\}$ and estimators from the range 3-25 for the RFC classifier. The optimal number of trees for the proposed framework for the RFC is to obtain 1000 trees and 10 estimators for each node for the AMD dataset.  The five-fold cross-validation is employed for multi-classification, which is illustrated in section \ref{subsec 3.2}
\\
To examine the performance of the MCGAEc method on the above-considered dataset, commonly used performance evaluation measures, including AUC (area under the receiver operating characteristic), Accuracy, Sensitivity, Specificity, F$_{1}$, and Matthews Correlation Coefficient (MCC) score is considered, which are described as follows:
\raggedbottom
\begin{equation}
    Accuracy = \frac{TP+TN}{TP+FP+TN+FN} \quad Sensitivity = \frac{TP}{TP+FN} 
\end{equation}
\begin{equation}
    Specificity = \frac{TN}{TN+FP} \quad F_{1} = \frac{2TP}{2TP+FP+FN}
\end{equation}
\begin{equation}
    MCC = \dfrac{TN \times TP -FN \times FP}{\sqrt{(TP+FP)(TP+FN)(TN+FP)(TN+FN)}}
\end{equation}
Where TP denotes true positive counts, FN denotes false negative counts, TN denotes true negative counts, and FP denotes false positive counts. For all the evaluation measures, higher values indicate better classification performance.   
\subsection{Effectiveness of the Proposed MCGAEc Model}
\begin{figure}[htbp]
\centering
\begin{minipage}{0.3\textwidth}
    \centering
    \includegraphics[width=3.1in]{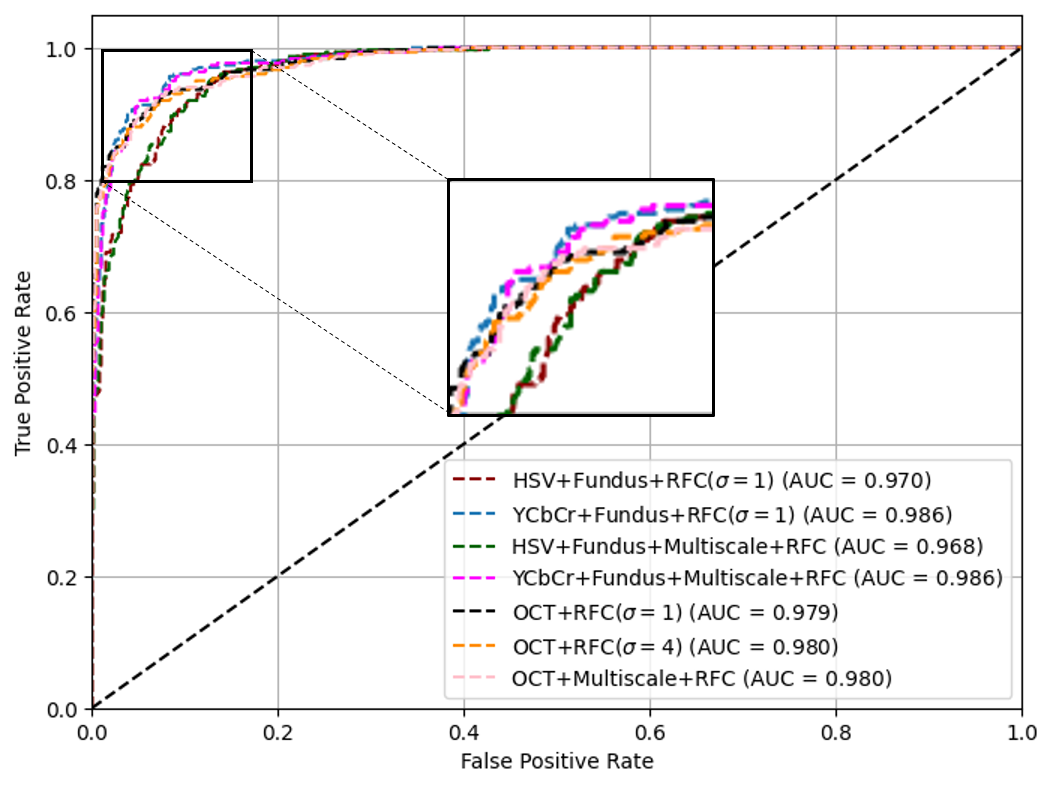}
    {(a)}
\end{minipage}
\\
\begin{minipage}{0.3\textwidth}
    \centering
    \includegraphics[width=3.1in]{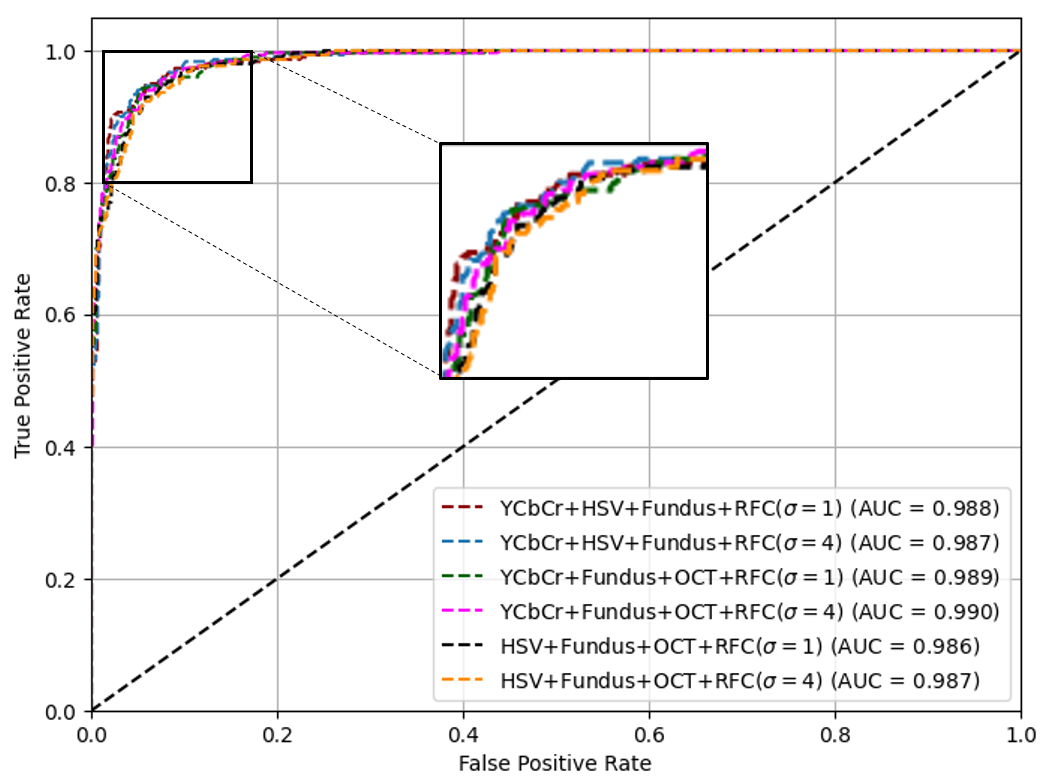}
    {(b)}
\end{minipage}
\\
\begin{minipage}{0.3\textwidth}
    \centering
    \includegraphics[width=3.1in]{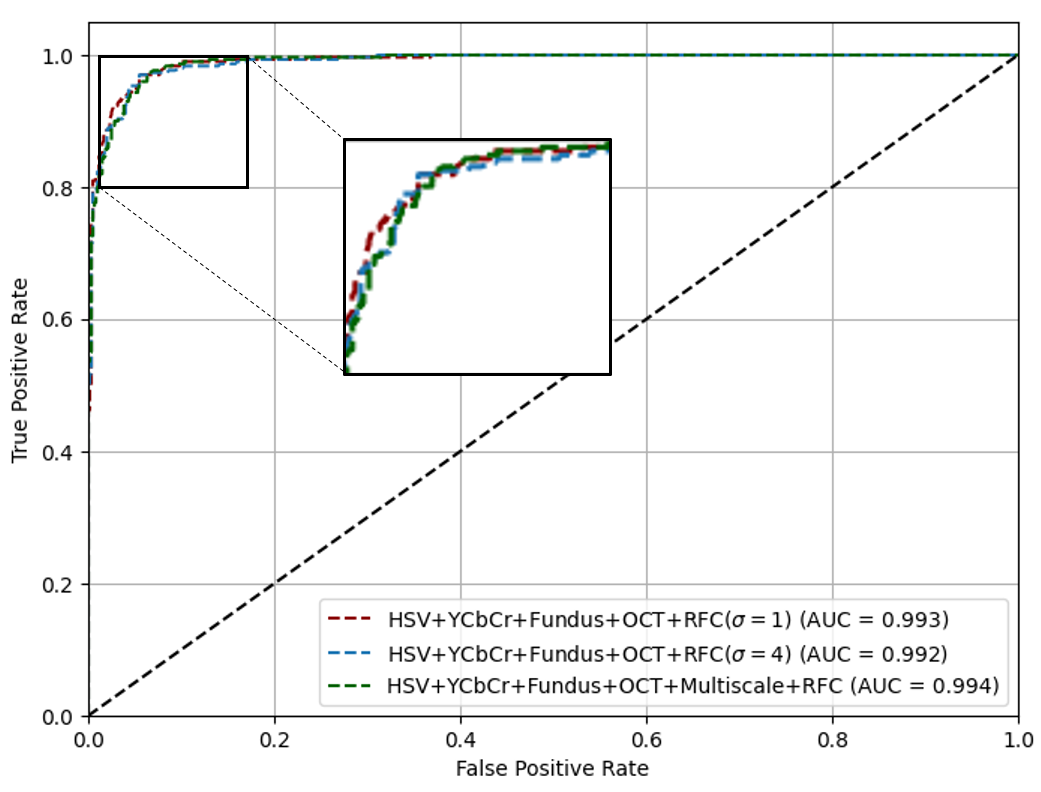}
    {(c)}
\end{minipage}
\caption{{ROC (Receiver operating characteristic) curves for the proposed method, (a) multiscale color space with single modality comparison, (b) fusion of the multiscale color spaces, (c) combination of multi-modality fundus and OCT images for the multi-class classification of AMD.}}
\label{fig 4}
\end{figure}

{\label{subsec 3.2}}
We first illustrate the comparison between the proposed MCGAEc model over the single modality by considering different color spaces with various scales. We have performed 5-fold cross-validation, and the experimental results are reported in Table \ref{table 1}.  When considering a single modality fundus image, we have performed the experiments on YCbCr and HSV color spaces by taking the lower and higher regularization levels to capture the features at different scales, as shown in Table \ref{table 1}. For OCT images, we have considered different scales for comparison with the proposed framework and other considered cases. The pre-trained VGG16 model is the backbone for the feature extraction from single modality and multi-modality with respect to different multi-scale color spaces.  
\newline
Fig. \ref{fig 4} demonstrates ROC curves for each considered case over the proposed framework for the AMD dataset. It can be observed from the ROC curve the proposed modality-specific multiscale color space embedding based on attention mechanism (MCGAEc model) for AMD classification is more adequate compared to other single modalities based on single color spaces. The proposed method achieves a higher AUC value of 0.994 compared to other single modalities with single color spaces, as demonstrated in Fig. \ref{fig 4}. Table \ref{table 1} signifies that the proposed MCGAEc model can improve the capability of AMD classification compared to the utilization of a single modality model. In the single modality, the fundus image, when transformed to YCbCr color space with regularization level $\sigma = 4$, has achieved a higher AUC of 0.990 compared to other single modality cases. Further, it achieves a higher F$_{1}$-score of 0.930 among the other single modality for classification. However, when the fundus image is transformed into HSV color space with multiscale, it achieves the lowest F$_{1}$ score compared with others. The proposed model has higher Accuracy, Sensitivity, Specificity, F$_{1}$, and MCC score compared to others demonstrated in Table \ref{table 1}. It indicates that when different color spaces are considered in multiscale and incorporated with the multi-modality Fundus and OCT images, the performance of the classifier is enhanced significantly, and it enables to capture of essential local and global discriminative features at different scales using attention mechanism.             
\begin{table}[!t]
\centering
    \caption{Experimental performance over the test set of the single modality model over the multi-modality model by considering different color spaces and scale space based on 5-fold cross-validation of the training set}
    \makebox[0.95 \textwidth][c]{       
          \resizebox{0.95 \textwidth}{!}{
    \begin{tabular}{@{\extracolsep{4pt}}l c c c c c c @{}}
    \hline
    \multirow{1}{*}{Methods} & \multicolumn{1}{@{}c@{}}{AUC} & \multicolumn{1}{@{}c@{}}{Accuracy} & \multicolumn{1}{@{}c@{}}{Sensitivity} & \multicolumn{1}{@{}c@{}}{Specificity} & \multicolumn{1}{@{}c@{}}{F$_{1}$} & \multicolumn{1}{@{}c@{}}{MCC}  \\
\midrule

HSV+Fundus+RFC$(\sigma=1)$  & 0.970 & 0.870 & 0.870  & 0.935  & 0.869  & 0.810 \\ 

YCbCr+Fundus+RFC$(\sigma=1)$  & 0.986 & 0.907 & 0.906  & 0.953  & 0.906 & 0.860 \\ 

HSV+Fundus+multiscale+RFC  & 0.968 & 0.857 & 0.856  & 0.928 & 0.855 & 0.803 \\

YCbCr+Fundus+multiscale+RFC  & 0.986 & 0.910  & 0.913 & 0.955 & 0.909   & 0.860\\

OCT+RFC$(\sigma =1)$  & 0.979 & 0.887  & 0.886 & 0.943  & 0.888  & 0.833\\

OCT+RFC$(\sigma =4)$  & 0.980 & 0.890 & 0.890  & 0.945  & 0.891  & 0.848\\

OCT+multiscale+RFC & 0.980 & 0.893 & 0.894 & 0.947 & 0.894  & 0.847\\

YCbCr+HSV+Fundus+RFC$(\sigma=1)$ & 0.988 & 0.910 & 0.910 & 0.955 & 0.909 & 0.865 \\

YCbCr+HSV+Fundus+RFC$(\sigma=4)$ & 0.987 & 0.907 & 0.907 & 0.953 & 0.906 & 0.879\\

\hline

YCbCr+Fundus+OCT+RFC$(\sigma =1)$ & 0.989 & 0.927 & 0.927 & 0.963 & 0.927 & 0.886\\

YCbCr+Fundus+OCT+RFC$(\sigma =4)$ & 0.990 & 0.930 & 0.932 & 0.965 & 0.930 & 0.895\\

HSV+Fundus+OCT+RFC$(\sigma =1)$ & 0.986 & 0.923 & 0.920 & 0.962 & 0.923 & 0.880\\

HSV+Fundus+OCT+RFC$(\sigma =4)$ & 0.987 & 0.916 & 0.917 & 0.958 & 0.916 & 0.861\\

HSV+YCbCr+Fundus+OCT+RFC$(\sigma =1)$ & 0.993 & 0.943 & 0.943 & 0.972 & 0.943 & 0.896\\

HSV+YCbCr+Fundus+OCT+RFC$(\sigma =4)$ & 0.992 & 0.937 & 0.932 & 0.968 & 0.937 & 0.892\\
\hline

\textcolor{blue}{\textbf{Proposed (MCGAEc)}} & \textcolor{blue}{\textbf{0.994}} & \textcolor{blue}{\textbf{0.947}} & \textcolor{blue}{\textbf{0.948}} & \textcolor{blue}{\textbf{0.973}} & \textcolor{blue}{\textbf{0.947}} & \textcolor{blue}{\textbf{0.907}}\\
\hline
    \end{tabular}}}
     \label{table 1}
\end{table}

\subsection{Quantitative Comparison with Existing Approaches}{\label{subsec 3.3}}

         In this section, we have compared the proposed model with the existing method over the considered AMD dataset, and the empirical study is demonstrated in Table \mbox{\ref{table 2}-\ref{table 3}}. For comparison purposes with the AMD dataset, the implementation of the SOTA methods is conducted.  The training results over the AMD dataset of the proposed framework are presented in Table \mbox{\ref{table 2}} and compared with the Yoo et al. \mbox{\cite{yoo2019possibility}}, which indicates that the proposed model is competent to extract the most discriminative features, which is essential for the classification of AMD in comparison to \mbox{\cite{yoo2019possibility}}. The proposed MCGAEc model enables to distinguish between false positive and true positive at distinct threshold levels indicated by the higher AUC value acquired by the proposed framework. Table \mbox{\ref{table 3}} represents the test result over the multi-modality fundus and OCT image dataset and is compared with the SOTA methods. The proposed model attained a higher AUC of 0.994 compared to other methods, which indicates the consistency of the proposed classification framework at different threshold levels except \mbox{\cite{fang2021multi}}, and it can be observed from the ROC curve illustrated in Fig. \mbox{\ref{fig 4}}. The presented method \mbox{\cite{fang2021multi}} acquires lower Specificity compared to our proposed method, which indicates the diagnosis of AMD in the false positive category.
        MCGAEc model performance is adequate in terms of quantitative comparison for the considered AMD dataset. The empirical analysis of the proposed MCGAEc model is adequate compared to the SOTA approaches illustrated in Table \mbox{\ref{table 3}} over the test set.             
\begin{table}[!t]
\centering
    \caption{Experimental performance of the proposed MCGAEc model compared with the state-of-the-art method \cite{yoo2019possibility} over the training set using five-fold cross-validation for the AMD dataset}
    \makebox[0.9 \textwidth][c]{       
          \resizebox{0.9 \textwidth}{!}{
    \begin{tabular}{@{\extracolsep{4pt}}l c c c c c @{}}
    \hline
    \multirow{1}{*}{Methods} & \multicolumn{1}{@{}c@{}}{AUC} & \multicolumn{1}{@{}c@{}}{Accuracy} & \multicolumn{1}{@{}c@{}}{Sensitivity} & \multicolumn{1}{@{}c@{}}{Specificity} & \multicolumn{1}{@{}c@{}}{F$_{1}$}  \\
\midrule
RF - OCT image alone & 0.906 &  0.826 & 0.833 & 0.816 & $-$ \\
RF - Fundus image alone  & 0.914 & 0.835 & 0.834 & 0.836 & $-$  \\
RF - OCT+Fundus image  & 0.969 & 0.905 & 0.910 & 0.896 & $-$  \\
RBM - OCT+Fundus image  & 0.940 & 0.865 & 0.860 & 0.875 & $-$ \\
DBN - OCT+Fundus image  & 0.956 & 0.889 & 0.880 &  0.905 & $-$  \\
\hline
\textcolor{blue}{\textbf{Proposed (MCGAEc)}} & \textcolor{blue}{\textbf{0.994}} & \textcolor{blue}{\textbf{0.944}} & \textcolor{blue}{\textbf{0.943}} & \textcolor{blue}{\textbf{0.972}} & \textcolor{blue}{\textbf{0.943}}\\
\hline
    \end{tabular}}}
     \label{table 2}
\end{table}
\begin{table}[!t]
\centering
    \caption{Experimental performance of the proposed MCGAEc model compared with the state-of-the-art method over the test set for the AMD dataset}
    \makebox[0.9 \textwidth][c]{       
          \resizebox{0.9 \textwidth}{!}{
    \begin{tabular}{@{\extracolsep{4pt}}l c c c c c @{}}
    \hline
    \multirow{1}{*}{Methods} & \multicolumn{1}{@{}c@{}}{AUC} & \multicolumn{1}{@{}c@{}}{Accuracy} & \multicolumn{1}{@{}c@{}}{Sensitivity} & \multicolumn{1}{@{}c@{}}{Specificity} & \multicolumn{1}{@{}c@{}}{F$_{1}$}  \\
\midrule

RF - OCT image alone \cite{yoo2019possibility} & 0.914 & 0.833 & 0.808 & 0.883 & $-$\\
RF - Fundus image alone \cite{yoo2019possibility} & 0.954 & 0.892 & 0.900 & 0.877 & $-$ \\
RF - OCT+Fundus image \cite{yoo2019possibility} & 0.981 & 0.946 & \textbf{0.955} & 0.927 & $-$  \\
RBM - OCT+Fundus image \cite{yoo2019possibility} & 0.976 & 0.931 & 0.942 & 0.910 & $-$ \\
DBN - OCT+Fundus image \cite{yoo2019possibility} & 0.961 & 0.892 & 0.888 &  0.913 & $-$  \\
Inception-v3 - Fundus image alone \cite{yoo2019possibility} & 0.958 & 0.892 & 0.900 & 0.877 & $-$\\
LASSO Regression - OCT+Fundus image \cite{yoo2019possibility} & 0.950 & 0.885 & 0.887 & 0.882 & $-$\\
ANN-Fundus segmentation \cite{yoo2019possibility} & 0.911 & 0.841 & 0.852 & 0.820 & $-$ \\

Yin et al. \cite{dai2021transmed} & 0.998 & 0.938 & 0.937 & 0.969 & 0.936\\

Xin et al. \cite{xing2022advit} & 0.991 & 0.925 & 0.925 & 0.963 & 0.924 \\

Fang et al. \cite{fang2021multi} & \textbf{0.996} & 0.945 & 0.941 & 0.951 & 0.946 \\

Aiyub et al. \cite{ali2024amdnet23}  & 0.989 & 0.945 & 0.939 & 0.952 & 0.946 \\
\hline
\textcolor{blue}{\textbf{Proposed (MCGAEc)}} & \textcolor{blue}{\textbf{0.994}} & \textcolor{blue}{\textbf{0.947}} & \textcolor{blue}{\textbf{0.948}} & \textcolor{blue}{\textbf{0.973}} & \textcolor{blue}{\textbf{0.947}} \\
\hline
    \end{tabular}}}
     \label{table 3}
\end{table}

\subsection{Discussion}{\label{subsec 3.4}}
This is the first experimental study, best to the knowledge of the authors to consider a multiscale color space for fusing distinct imaging modalities for the classification of AMD disorder. In this study, we proposed an MCGAEc model that considers the fundus and OCT images at multiscale color spaces simultaneously for the diagnosis of AMD. Fundus imaging modality characterizes information on the region of the drusen (AMD). OCT provides subsurface cross-section imaging, providing information about the different layers of the retina. Such information is complimentary to the structural information of the vasculature on the surface as obtained from fundus images. OCT imaging modality is correlated with the subsurface of retinal layers and intra-retinal fluid lesions. The thickness of retinal layers, which is influenced by choroidal neovascularization, is investigated through the OCT, and the fundus image is competent to apprehend the evolution in the size of the drusen. However, the fundus image is insufficient in identifying choroidal neovascularization rigorously \cite{mokwa2013grading}. Whereas OCT is not able to identify the transitions in the drusen and retinal pigment epithelium \cite{castillo2015optical}. The fundus and OCT imaging modalities provide complementary information on the retina. Early stages-based techniques utilized fundus or OCT for diagnosis of AMD based on deep learning models \mbox{\cite{fu2018disc,li2016integrating,lee2017deep, kruper2024convolutional,das2023net}}. This work focuses on investigating the crucial feature information extracted from Fundus and OCT images, which are integrated with the proposed MCGAEc method to capture the surface and subsurface retinal information for the diagnosis of AMD and glaucoma. We have assimilated multiscale color space to diagnose complex retinal diseases, which helps to capture dissimilar features from OCT and Fundus simultaneously at distinct color spaces with multiple scales. In the proposed approach, we attempt to combine distinct feature properties preserved in the various color spaces and if it is captured at a finer scale so there is a possibility of the presence of that particular feature at multiple scales. Therefore, we have considered YCbCr and HSV color spaces at different scales and integrated them with the pre-trained VGG16 model to extract the crucial feature for diagnosis purposes. On the other hand, OCT images are considered on different scales to capture the subsurface retinal layer information for the diagnosis of choroidal neovascularization. The attention mechanism is incorporated to extract global feature representation and integrated with the local feature information, followed by the ensembling of each classifier at the feature fusion module, and a random forest classifier is utilized for the classification of various stages of AMD. To show the significance of the modality-specific multiscale color space embedding, the experiments are also performed over a single modality based on different scales of color spaces, and results are illustrated in Table \ref{table 1}. It can be observed from Table \ref{table 1} that when the fundus image is utilized and transformed into YCbCr color space at multiscale, it achieves 0.910 Accuracy for the AMD dataset compared to other single modalities with one color space transformation. However, the experimental performance is similar for single modality fundus images when distinct color spaces are fused together. On the other hand, when modality-specific multiscale color space embedding strategy is assessed, then the performance of the MCGAEc model is elevated and achieved 0.947 Accuracy for the AMD dataset. Further, we have performed a comparative study demonstrated in Table \ref{table 2} and \ref{table 3}, which shows the efficacy of the proposed MCGAEc method over the SOTA methods. The proposed study can be integrated into the clinical setting to help ophthalmologists with the diagnosis of the retinal disorder and, based on the observation, can predict the retinal disease.
\section{Conclusion}{\label{sec 4}}
We have presented a multi-modality MCGAEc deep learning model that assimilates the feature representations in various color spaces and emphasizes the vital discriminative features by multiscale mechanism. Besides, the global feature representation is incorporated into the proposed model by employing the attention mechanism at each path of the ensemble classifier. To analyze the behavior of the proposed model, extensive experiments were accomplished over the publicly available multi-modality AMD dataset and compared with the existing approach, which indicates the effectiveness of the proposed model. Additionally, the significance of the proposed method has been verified by performing experiments on a single modality with distinct color spaces, and the proposed MCGAEc model achieves higher evaluation measures compared to the considered cases. Moreover, we successfully incorporated ROI-specific mechanisms to learn essential features from the multi-modality imaging techniques used to diagnose retinal disorders and localize the affected region.
\section*{Declarations}
\textbf{Conflict of interest} The authors declare that they have no conflict of
interests.

\section*{Acknowledgement}
This work is supported by the Grant of SERB, Government of India (Ref. No: SPG/2022/000045 ) and IIT Kharagpur AI4ICPS I Hub  Foundation (Ref No:  IIT/SRIC/MM/ZDP/2023-2024/156).
\bibliographystyle{splncs04}
\bibliography{References}
\end{document}